\documentclass[12pt,a4paper]{article}
\usepackage{latexsym}
\usepackage{amsmath}
\usepackage{amsfonts}
\usepackage{amsbsy}
\usepackage{amssymb}
\usepackage{graphicx}
\usepackage{mathrsfs}

\def \R{{\mathbb R}}
\def \C{{\mathbb C}}

\def \N{\mathbb{N}}

\def \mm{\boldsymbol{\boldsymbol{\mu}}}
\def \KMS{\omega_{\beta}}
\def \BHB{\omega_{hb}}

\def \Cyl{{\mathcal C}}
\def \D{{\mathscr D}}

\def \F{{\mathcal F}}
\def \Ga{\mathcal{G}}

\def \L{\Lambda}

\def \O{\mathcal{O}}

\def \S{{\mathcal S}}

\def \T{{\mathcal T}}
\def \a{\alpha}
\def \b{\beta}

\def \d{\delta}

\def \l{\lambda}

\def \s{\sigma}

\def \e{\varepsilon}
\def \o{\omega}

\def \t{\tau}
\def \th{\theta}
\def \z{\zeta}
\def \n{\nu}
\def \m{\mu}

\def \bpsi{\bar{\psi}}
\def \dr{\dot{r}}
\def \ds{\dot{s}}
\def \dt{\dot{t}}

\def \del{\partial}

\def \supp{\mbox{supp }}

\newtheorem{Theorem}{Theorem}[section]

\newtheorem{Lemma}{Lemma}[section]
\newtheorem{Definition}{Definition}[section]

\numberwithin{equation}{section}

\begin{document}

\title{The Hot Bang state of massless fermions\\[20pt]}

\author{Benjamin Bahr\\ Institute for Theoretical Physics\\ University of G\"ottingen\\[20pt]}

\maketitle

\abstract{\noindent
%According to the characterization of local
%thermal equilibrium (LTE) states in Local Quantum Physics proposed
%by Buchholz, Ojima and Roos, the thermal observables and functions
%are computed for the model of massless, free fermions on Minkowski
%space. Furthermore, evolution equations for the phase-space
%density are derived, that are more restrictive than in the bosonic
%case. In the end, an example for an LTE state is given and its
%positivity is proven.
In 2002, a method has been proposed by Buchholz et al. in the
context of Local Quantum Physics, to characterize states that are
locally in thermodynamic equilibrium. It could be shown for the
model of massless bosons that these states exhibit quite
interesting properties. The mean phase-space density satisfies a
transport equation, and many of these states break time reversal
symmetry. Moreover, an explicit example of such a state, called
the Hot Bang state, could be found, which models the future of a
temperature singularity. However, although the general results
carry over to the fermionic case easily, the proof of existence of an
analogue of the Hot Bang state is not quite that straightforward. The proof will be given 
in this paper. Moreover, we will
discuss some of the mathematical subtleties which arise in the
fermionic
case.\\[20pt]}

\section{Introduction}

\noindent In the framework of Local Quantum Physics, states that
describe global thermodynamic equilibrium are well-known and are
characterized by the KMS-condition \cite{HAAG}. For many models
the KMS states are known and relatively simple to write down
explicitly. On the other hand it is a nontrivial problem to obtain
states that describe systems which are only locally in
thermodynamic equilibrium, such as hydrodynamic flows and heat
transfers. This is due to the fact that such a state should
describe a system that thermalizes on a small, but not on a large
scale, thus behaving significantly differently on these two
scales.

 In \cite{BOR02},
Buchholz, Ojima and Roos proposed a method to make this idea
mathematically precise. They tried to characterize states that
describe situations of local thermodynamic equilibrium. This
characterization uses the KMS states as a tool to locally compare
a given state with global equilibrium states in order to assign
thermodynamic properties to that state. This comparison is
point-dependent and thus delivers a way to describe notions of
temperature or entropy that may vary from point to point.

This method has been applied to the model of massless, free bosons
on $\R^4$, which has led to interesting results. Firstly, the
microscopic dynamics induces a macroscopic transport equation for
the phase-space density. Secondly, one finds that states that have
a thermodynamic interpretation in a sufficiently large region
break time-reversal symmetry, thus implementing a thermodynamic
arrow of time \cite{BU03, BOR02}. Thirdly, an example for a local equilibrium
state is given by the so-called Hot-Bang state that describes the
effects of a heat explosion at some point.

Most of the results stated above carry over to the fermionic case
easily. They will only be mentioned in short in this paper. What
poses a problem is to establish the existence of an analogy of the
Hot-Bang state. The proof of the existence of such a state will
cover the main part of this work.

\section{Massless free fermions}
\noindent The setting for the analysis will be the CAR-Algebra of
massless free fermions. In the massless case, the Dirac equation
decomposes into two independent equations, called Weyl equations.
They describe the left-handed and the right-handed part of the
Fermion separately. Thus, we will consider the smeared-out fields
$\psi(f)$ and $\bpsi(f)$,  where $f$ is a smooth function with
compact support and takes values in $\C^2$. These $\psi(f)$ and
$\bpsi(f)$ create a $C^*$-algebra $\F$  subject to the relations:
\begin{align}\label{Antikommutatorzwei1}
\Big\{\psi(f),\bpsi(g)\Big\}\;&=\;2\pi\int\,dp\;
\d(p^2)\e(p_0)\tilde{f}(p)^T\,p_M\,\tilde{g}(-p)\,\cdot\,1\\[5pt]\label{Antikommutatorzwei2}
\Big\{\psi(f),\psi(g)\Big\}\;&=\;\Big\{\bpsi(f),\bpsi(g)\Big\}\;=\;0,
\end{align}

\noindent (where $\e(x)=\th(x)-\th(-x)$ is the sign distribution) and          %\cite{BRII}, \cite{HAAG}, \cite{BOR96}

\begin{align}\label{Diraczwei}
\psi(-i(\del^M)^Tf)=\bpsi(i\del^Mf)=0.
\end{align}

\noindent Here, for a vector $a\in\R^4$ the $2\times 2$-matrices
$a_M$ and $a^M$ are defined by

\begin{align*}%\label{Gl:AMundAM}
a_M\doteq\left(\begin{array}{cc}a^0+a^3&a^1-ia^2\\a^1+ia^2&a^0-a^3\end{array}\right)\qquad\mbox{und}\qquad
a^M\doteq\left(\begin{array}{cc}a^0-a^3&-a^1+ia^2\\-a^1-ia^2&a^0+a^3\end{array}\right).
\end{align*}

\noindent Furthermore, the $*$-relation is given by

\begin{align}\label{Sternrelationenzwei}
\psi(f)^*=\bpsi(\bar{f})\qquad\bpsi(f)^*=\psi(\bar{f}),
\end{align}

\noindent where $\bar{f}$ is the componentwise complex conjugate
function of $f$. The double covering of the Poincar\'e group, the
elements of which consist of pairs $(A, a)$ with $A\in SL(2,\C)$,
$a\in\R^4$, acts on $\F$ via
\begin{align}\label{Poincarezwei1}
\a_{(A,a)}\psi(f)&=\psi((A^T)^{-1}f_{(\L,a)})\\[5pt]\label{Poincarezwei2}
\a_{(A,a)}\bpsi(f)&=\bpsi((A^{\dag})^{-1}f_{(\L,a)})
\end{align}

\noindent with $f_{(\L,a)}(x)=f(\L^{-1}(x-a))$, the Lorentz
transform $\L$ being the corresponding one to $A\in SL(2,\C)$. The
global gauge group $U(1)$ acts on $\F$ by
\begin{align}\label{Gl:Eichgruppezwei}
\a_{\varphi}\psi(f)\,\doteq\,e^{i\varphi}\psi(f)\,\qquad\qquad
\a_{\varphi}\bpsi(f)\,\doteq\,e^{-i\varphi}\bpsi(f).
\end{align}

\noindent In the case of $\F$, the KMS-states and their properties
are known \cite{BRII}. Since the theory is massless and free, the
global equilibrium situations need to be labelled by inverse
temperature $|\b|>0$, but not by chemical potential. Furthermore,
every KMS state determines the rest system, with respect to which
it is in equilibrium, due to the fact that Lorentz symmetry is
spontaneously broken in KMS states. A rest system is uniquely
defined by a future directed, timelike unit vector $e$. We combine
these two parameters to a vector in the forward lightcone, which
we denote by $\b=|\b| e$. We will consider gauge-invariant
KMS states only, and for each temperature-vector $\b\in V^+$ there
is a unique gauge-invariant KMS-state $\KMS$ \cite{BAB04}. All these states are
quasifree and thus completely determined by their
two-point-function, which is given by
\begin{align}\label{Gl:KMS1}
&\KMS\Big(\bpsi(f)\,\psi(g)\Big)\;=\;2\pi\int_{\R^4}dp\,\d(p^2)\,\e(p_0)
\,\frac{\tilde{g}^T(p)\,p_M\,\tilde{f}(-p)}{1+e^{-(\b,p)}}\\[5pt]\label{Gl:KMS2}
&\KMS\Big(\psi(f)\,\psi(g)\Big)\;
=\;\KMS\Big(\bpsi(f)\,\bpsi(g)\Big)\;=\;0.
\end{align}
%\KMS\Big(\psi(f)\Big)\;=\;\KMS\Big(\bpsi(f)\Big)\;=\;

\noindent A special case of this
is the so-called vacuum state $\o_{\infty}$. It is given by
(\ref{Gl:KMS1}) and (\ref{Gl:KMS2}), where $\b$ tends to timelike
infinity, i.e. one has:
\begin{align}\label{Gl:Vak1}
&\o_{\infty}\Big(\bpsi(f)\,\psi(g)\Big)\;=\;2\pi\int_{\R^4}dp\,\d(p^2)\,\th(p_0)
\,{\tilde{g}^T(p)\,p_M\,\tilde{f}(-p)}.
\end{align}

\section{Local equilibrium states}

\noindent Let $B$ be a compact subset of $V^+$ and $d\rho$ be a
normalized measure on $B$. Due to (\ref{Gl:KMS1}) the function
$\b\mapsto\KMS(A)$ is continuous and thus one can form the
statistical mixtures of KMS-states:
\begin{align}\label{Gl:Mixture}
\o_{B}\,=\,\int_Bd\rho(\b)\,\KMS.
\end{align}

\noindent The mixtures for all $B$ and all $d\rho$ form the set
$\Cyl$ of so-called reference
states.\\[5pt]

\noindent With the reference states at hand, we are able to
compare a given state with them at a point in order to analyze the
thermal properties of that state at that point. We do this by
testing the states on a set of observables that correspond to
measurements of thermal properties at a single point. It is
obvious that the observables in $\F$ cannot be used for this,
since they consist of the field smeared over a finite region in
spacetime. To proceed, we need to go
over to idealized observables, that exist in the sense of forms.\\

\noindent Let $\mm=(\m_1\cdots\m_m)$ be a multi-index. We define
the following 'observables':
\begin{align}\nonumber
\l^{\mm\n}(x)\,&\doteq\,\eth^{\mm}\,:\bpsi_{\dr}(x)\s^{\n,\dr
s}\psi_s(x):\\[5pt]\label{Gl:Lambdas1}
&\doteq\,\lim_{\begin{array}{c}\scriptstyle\z\to0\\\scriptstyle\z^2<0\end{array}}\;
\del^{\mm}_{\z}\;:\bpsi_{\dr}(x+\z)\s^{\n,\dr s}\psi_s(x-\z):
\end{align}

\noindent where the normal ordering is performed with respect to
the vacuum state $\o_{\infty}$. The $\l^{\mm\n}(x)$ are called
thermal observables at $x\in\R^4$ and correspond to measurements
at $x$. They are idealizations: One cannot expect the expression
$\o\big(\l^{\mm\n}(x)\big)$ to make sense for arbitrary states.
This idealization is needed in order to distinguish thermal
properties at different but arbitrary close points. We will only
consider states in which the limit (\ref{Gl:Lambdas1})
exists.\\

\noindent For $x\in\R^4$, the linear span of the $\l^{\mm\n}(x)$
(for all $\mm$) will be denoted by $\S_x$. The elements in $\S_x$
transform the way their tensor indices indicate:
\begin{align*}
\a_{(S,a)}\l^{\mm\n}(x)\;=\;\L_{\m_1'}{}^{\m_1}\cdots\L_{\m_m'}{}^{\m_m}\L_{\n'}{}^{\n}\l^{\mm'\n'}(\L
x+a).
\end{align*}

\noindent Thus, the spaces $\S_x$ are transformed into each other
by the action of the automorphisms via $\a_{y-x}\S_x=\S_y$. These
thermal observables and the reference states are used to
characterize local equilibrium states:

\begin{Definition}\label{Def:Esso}
Let $\o$ be a state over $\F$ and $\O\subset\R^4$ be open. The
state $\o$ is called $\S_{\O}$-thermal, if the following
conditions hold:
\begin{itemize}
\item[(i)] For every $x\in\O$ there is a reference state
$\o_{B_x}\in\Cyl$ such that
$\o\big(\l^{\mm\n}(x)\big)=\o_{B_x}\big(\l^{\mm\n}(x)\big)$ for
all $\l^{\mm\n}(x)\in\S_x$.

\item[(ii)] For every compact subset $U\subset\O$ there is a
compact subset $B\subset V^+$ such that the regions $B_x$ for all
$x\subset U$ lie all in $B$.
\end{itemize}
\end{Definition}

\noindent One would think of an $\S_{\O}$-thermal state as one
being close to global equilibrium at every point $x\in\O$, because
at this point it coincides with some global equilibrium state on
the set of thermal observables.\\

\noindent For an element $\l^{\mm\n}(x)$ the corresponding
function
\begin{align}\label{Gl:ThermalFunction}
V^+\,\ni\,\b\,\longmapsto\,\KMS\big(\l^{\mm\n}(x)\big)\,\doteq\,L^{\mm\n}(\b)
\end{align}

\noindent is called thermal function. By straightforward
calculation one shows that
\begin{align}\label{FormL}
L^{\mm\n}(\b)=\KMS(\l^{\mm\n}(x))=c_m\left(\del_{\b}^{\mm\n}\frac{1}{(\b,\b)}\right),
\end{align}

\noindent with $m=\deg \mm$ and
\begin{align}\label{cm}
c_m=\left\{\begin{array}{cl}\frac{i\,\pi^{m+1}\;(2^{2m+2}-2^{m+1})}{(m+3)!}\;(-1)^{\frac{m+3}{2}}\;B_{\frac{m+3}{2}}\quad
&\mbox{for odd $m$}\\[15pt]
0\qquad &\mbox{for even $m$}\end{array}\right\},
\end{align}

\noindent where the $B_n$ are the Bernoulli numbers. Since the
KMS-states $\KMS$ are translation-invariant, the value of
$L^{\mm\n}(\b)$ does not depend on $x$. It indicates what
expectation value the thermal observables have in the global
equilibrium states. It shows why the choice of (\ref{Gl:Lambdas1})
as thermal observables is sensible: By thermodynamic
considerations \cite{BOR02, DIX} one knows what value intensive
thermal properties such as energy density, entropy current density
and phase space density should have in the global equilibrium
states. The thermal energy-momentum tensor in a system of
massless, free fermions being in a state of constant temperature
$\b\in V^+$, for example, has the form:
\begin{align}\label{Gl:Energyexpectation}
E^{\m\n}(\b)\,=\,\frac{\pi^2}{60}\left(\frac{4\b^{\m}\b^{\n}}{(\b,\b)^3}-\frac{\eta^{\m\n}}{(\b,\b)^2}\right)
\end{align}

\noindent at every point $x\in\R^4$. In fact, the thermal
observable
\begin{align}\label{Gl:EnergyMomentumTensor}
:\th^{\m\n}(x):\;\doteq\;\frac{1}{2i}(\l^{\m\n}(x)+\l^{\n\m}(x))
\end{align}

\noindent is not only the normal ordered, symmetrized
energy-momentum tensor of the free, massless Dirac field, but we
also have $\KMS(:\th^{\m\n}(x):)=E^{\m\n}(\b)$, as one can see by
(\ref{FormL}). So in the $\S_x$ there is an observable for the
thermal energy density at $x\in\R^4$. In fact, the $\S_x$ contain
enough elements to approximate all important thermal properties of
a system, as will be shown in the following. This situation is
similar to the bosonic case.

\section{Admissible macroobservables and transport equations}

\noindent Since $\b\mapsto(\b,\b)^{-1}$ solves the wave equation
on $V^+$, we see by (\ref{FormL}) that all thermal functions do so
too: $\Box_{\b} L^{\mm\n}(\b)=0$. In fact, if one introduces a
family of seminorms on the space of continuous functions on $V^+$
via
\begin{align}\label{Gl:Seminorms}
\|\Xi\|_B\,\doteq\,\sup_{\b\in B}\,|\Xi(\b)|
\end{align}

\noindent where $B\subset V^+$ is compact and indexes this family,
then the set of smooth solutions of the wave equation on $V^+$
becomes a pre-Frech\`et space, call it $\Ga$. One can show
\cite{BU03, BAB04} that with respect to the seminorms
(\ref{Gl:Seminorms}) the space of all thermal functions $\b\mapsto
L^{\mm\n}(\b)$ is dense in $\Ga$. Thus in the spaces of thermal
observables there are elements that approximate other idealized
observables whose thermal functions are smooth solutions of the
wave equation on $V^+$. From \cite{DIX} and \cite{BOR02, BAB04} it
is known that for massless, free fermions in an equilibrium state
with inverse temperature $\b\in V^+$ the entropy-current density
$S^{\m}$ is given by
$S^{\m}(\b)=\frac{\pi^2}{15}\frac{\b^{\m}}{(\b,\b)}$. Furthermore,
the phase-space density of such a system is
$N_p(\b)=(2\pi)^{-3}(1+e^{(\b,p)})^{-1}$ (i.e. both are constant
in the space-time variable). Both are solutions of the
wave-equation on $V^+$ (in $\b$). So, given any compact set on
$B\subset V^+$, one can find elements in $\S_x$ that approximate
the observables entropy-current density and phase-space density on
all $\o_{B'}$ with $B'\subset B$ arbitrarily well. Condition (ii)
in definition \ref{Def:Esso} guarantees that all $\S_{\O}$-thermal
states are continuous with respect to the seminorms
(\ref{Gl:Seminorms}), so one can assign an expectation value of
$N_p$ or $S^{\m}$ to such a state $\o$ at every point $x\in\O$ by
the following rule: For $x\in\O$ let $\phi_n(x)$ be a sequence of
elements in $S_x$ whose thermal functions $\Phi_n(\b)$ tend to
$N_p(\b)$. Then define the phase-space density of the system in
the state $\o$ at $x\in\O$, $p\in\del\overline{V}^+$ to be
\begin{align}\label{Gl:approx}
\o(N_p)(x)\,\doteq\,\lim_{n\to\infty}\o\big(\phi_n(x)\big).
\end{align}

\noindent By similar constructions, one can define the expectation
values of $S^{\m}$ or other desired properties, such as free
energy or Gibbs-Potential, in $\o$ at every $x\in\O$. One can show
\cite{BU03} that by this procedure every element $\Xi$ in $\Ga$
determines an observable (again called $\Xi$) commuting with all
elements in $\F$. Let $\o$ be an $\S_{\O}$-thermal state, then one
can assign an expectation value of $\Xi$ in $\o$ to every $x\in\O$
by
\begin{align}\label{Gl:DefXI}
\o(\Xi)(x)\,\doteq\,\int_{B_x}d\rho_x(\b)\,\Xi(\b).
\end{align}

\noindent This generalizes (\ref{Gl:approx}). The $\Xi$ are called
admissible macro-observables. So $S^{\m}$, $N_p$ and $E^{\m\n}$
thus are such macroobservables interpreted as intensive thermal
properties of the system, whose mean values in global equilibrium
states are determined by their thermal function $\b\mapsto
\Xi(\b)$. By (\ref{Gl:DefXI}) one can assign such a value to every
point in $\O$ to a system being in an $\S_{\O}$-thermal state
$\o$.\\

\noindent Condition (ii) in definition \ref{Def:Esso} assures the thus
constructed functions $x\mapsto \o(\Xi)(x)$ to be differentiable
in $x$ in the sense of distributions. Moreover, one would think of
this function as the point-dependent mean value of the admissible
macroobservable $\Xi$ in the $\S_{\O}$-thermal state $\o$. So
thermal properties such as energy density or entropy current
density can vary from point to point. For instance, one would
interpret the function
\begin{align}\label{Gl:PSDe}
\O\times\del\overline{V}^+\;\ni\;(x,p)\,\longmapsto\,\o(N_p)(x)
\end{align}

\noindent to be the mean phase-space density of the system in the
state $\o$. As one can show \cite{BU03, BAB04}, the Weyl equations
(\ref{Diraczwei}) determine an evolution equation for
(\ref{Gl:PSDe}). Let $p\in\del\overline{V}^+$ be a positive,
lightlike vector, then one finds, for example, that
\begin{align}\label{Gl:PSD}
p_{\m}\del^{\m}\,\o(N_p)(x)\,=\,0
\end{align}

\noindent (where the derivative is to be taken with respect to
$x$). This is the collisionless, free Boltzmann equation. So a
transport equation for locally thermal states can be derived from
first principles and does not need to be imposed on the system.
Again, this feature quite mimics the situation in the case of
massless bosons.

\section{The Hot-Bang state}

\noindent In the massless bosonic case, there is a special state
$\BHB^{Bos}$, called the Hot Bang state, whose features have been
exhibited in \cite{BU03}. It describes the effects of a heat
explosion at the origin of Minkowski space, i.e. $\BHB^{Bos}$ is an
$\S_{V^+}$-thermal state that describes a system with diverging
temperature on the boundary of the forward lightcone. It is (up to
reflections and translations) the only $\S_{\O}$-thermal state
that has a KMS-state as reference state at each point in $\O$.
That is, the state describes a system with locally sharp
temperature.

One would hope an analogous state $\BHB$ to exist in the massless
fermionic case, too. It can be shown \cite{BAB04} that the
condition of local sharpness of $\b$ determines the two-point
function of such a state to be

\begin{align}\label{Gl:BHB}
%\BHB\big(\bpsi(\bar{f})\psi(g)\big)\;=\;(2\pi)^{-3}\int_{\R^4}dp\,\d(p^2)\,\e(p_0)\,\int\,dx\,dy\,g^T(x)p_M
%\overline{f(y)}\,\frac{e^{i(p,x-y)}}{1+e^{\lambda(x+y,p)}}\\[10pt]
&\BHB\Big(\bpsi_{\dr}(x)\psi_s(y)\Big)\;=\;(2\pi)^{-3}\int_{\R^4}dp\,\d(p^2)\,\e(p_0)\,p_{s\dr}
\,\frac{e^{i(p,x-y)}}{1+e^{\lambda(x+y,p)}}\\[10pt]\label{Gl:BHB1}
&\BHB\Big(\psi_{r}(x)\psi_s(y)\Big)\;=\;\BHB\Big(\bpsi_{\dr}(x)\bpsi_{\ds}(y)\Big)\;=\;0.\vspace{5pt}
\end{align}

\noindent It is straightforward to show that (\ref{Gl:BHB}) and
(\ref{Gl:BHB1}) define a linear, quasifree, gauge-invariant
functional on a dense subset of the CAR-Algebra $\F$. What is less
clear is whether this functional is a state, that is if
$\BHB(A^*A)\geq0$ for all $A\in\F$. So it is not clear, whether an
analogue to the bosonic Hot-Bang-state exists. There, the proof of
positivity for the corresponding functional is quite short and straightforward.
The proof for the above functional to be positive, on the other
hand, will cover the rest of this chapter.\\

\noindent What we will show in this chapter is that (\ref{Gl:BHB})
is positive for $x,y\in V^+$, that is $\BHB(A^*A)\geq 0$ for all
$A\in\F(V^+)$. Here $\F(V^+)$ denotes the sub-$C^*$-algebra of
$\F$ that is generated by all $\psi(f)$ and $\bpsi(f)$ with $\supp
f\subset V^+$. This again is similar to the bosonic case, where
the state $\BHB^{Bos}$ exhibits thermal properties on $V^+$ only.
In fact, neither in the bosonic nor in the fermionic case can this
region be enlarged, by quite general arguments \cite{BU03,
BAB04}.\\

\noindent By an argument in \cite{BRII}, one does not need to test
this condition for all $A\in\F(V^+)$, but only on $\psi(f)$ and
$\bpsi(f)$, $f\in\D(V^+,\C^2)$, because the functional is
quasifree. So we only need to show that
\begin{align*}%\label{Gl:Positivbedingung}
&\BHB\big(\bpsi(\bar{f})\psi(f)\big)\,\geq\,0\\[5pt]
&\BHB\big(\psi(f)\bpsi(\bar{f})\big)\,\geq\,0
\end{align*}

%Hab dich lieb!!!

\noindent for all $f\in\D(V^+,\C^2)$ to establish the result.\\

\noindent First of all, we consider some functional analytic
arguments.

\begin{Lemma}\label{Lem:Fvonz}
Let $f\in\D(V^+,\C^2)$. Let $\C_+\doteq\{z\in\C\,|\,\text{Im }
z\,>0\}$, then
\begin{align}\label{Gl:DefinitionvonF}
F(z)\doteq\int_{\R^3}\frac{d^3p}{2|\vec{p}|}\tilde{f}^T(zp')\,p'_M\,\overline{\tilde{f}(\overline{z}^{-1}p')}
\end{align}

\noindent exists for $z\in\overline{\C_+}\backslash\{0\}$ and is
continuous in $z$. Furthermore, $z\mapsto F(z)$ is holomorphic on
$\C_+$.
\end{Lemma}

\noindent\textbf{Proof:} First consider the complex Fourier
transform of $f$, with $\z\in\C^4$, which is an entire analytic
function in $\C^4$:
\begin{align*}
\tilde{f}(\z)=\frac{1}{(2\pi)^2}\int_{\R^4}dx\,e^{i(\z,x)}f(x).
\end{align*}

\noindent Because of $\supp f\subset V^+$, the theorem of
Paley-Wiener can be written down like this:

\begin{align}\label{Gl:WichtigeAbschaetzung}
\big|\,\tilde{f}(zp')\,\big|\,\leq\,C_N\frac{e^{-\d\,|\vec{p}|
\,\text{Im }z}}{(1+|z|\,|\vec{p}|)^N}\qquad\qquad\mbox{for all
}z\in\overline{\C_+}.
\end{align}

\noindent For fixed $p'=(|\vec{p}|,\vec{p})\in\del\overline{V}^+$
the integrand
\begin{align*}
z\,\longmapsto&\,\frac{1}{2|\vec{p}|}\tilde{f}^T(zp')\,p'_M\,\overline{\tilde{f}(\overline{z}^{-1}p')}\\[5pt]
=&\,\frac{1}{2|\vec{p}|}\int_{\R^8}dx\,dy\;{f}^T(x)\,p'_M\,\overline{{f}(y)}\,e^{i(p',
zx-z^{-1}y)}.
\end{align*}

\noindent is holomorphic on $\C\backslash\{0\}$, since $f$ has
compact support. By the explicit form of $p'_M$ one sees that
every one of its components is bounded by $2|\vec{p}|$. Using this
and the estimate (\ref{Gl:WichtigeAbschaetzung}), one sees that
the integrand is dominated by

\begin{align}\label{Gl:Abschaetzung2}
\left|\frac{1}{2|\vec{p}|}\tilde{f}^T(zp')\,p'_M\,\overline{\tilde{f}(\overline{z}^{-1}p')}\right|\;
\leq\;\frac{2C_N^2}{(1+|\vec{p}|^2)^N}
\end{align}

\noindent for $z\in\overline{\C_+}\backslash\{0\}$. Thus, if $z$
varies in some compact subset of $\C_+$, the integrand is
uniformly bounded by an integrable function of $\vec{p}$. Hence
the integral exists and is holomorphic in $\C_+$. Furthermore, if
$\{z_n\}_{n\in\N}$ is a sequence in $\C_+$ converging to
$r\in\R\backslash\{0\}$, we may interchange integration and limit
and get
\begin{align}
\lim_{z\to
r}\;\int_{\R^3}\frac{d^3p}{2|\vec{p}|}\tilde{f}^T(zp')\,p'_M\,\overline{\tilde{f}(\overline{z}^{-1}p')}
\;=\;\;\int_{\R^3}\frac{d^3p}{2|\vec{p}|}\tilde{f}^T(rp')\,p'_M\,\overline{\tilde{f}(r^{-1}p')},
\end{align}

\noindent which was the actual claim.

\begin{Theorem}\label{Satzzwei}
Let $f\in\D(V^+,\C^2)$. Then the function
\begin{align}\label{Gl:DefinitionvonL}
[0,\pi]\,\ni\,\phi\,\longmapsto\,
L(\phi)=\int_{\R^3}\frac{d^3p}{2|\vec{p}|}\tilde{f}^T(e^{i\phi}p')\,p'_M\,\overline{\tilde{f}(e^{i\phi}p')}\,\in\,\R
\end{align}

\noindent is either identically zero or logarithmically convex and
positive. Furthermore it is continuous on $[0,\pi]$ and smooth on
$(0,\pi)$.
\end{Theorem}

\noindent\textbf{Proof:} The claim about the continuity and
smoothness is evident from Lemma (\ref{Lem:Fvonz}) and
$L(\phi)=F(e^{i\phi})$. By the Cauchy-Schwartz-inequality and
scaling we get for
$z=re^{i\phi}\in\overline{\C_+}\backslash\{0\}$:
\begin{align*}
|F(z)|^2&\leq\;\int_{\R^3}\frac{d^3p}{2|\vec{p}|}\,\tilde{f}^T
(re^{i\phi}p')p'_M\overline{\tilde{f}(re^{i\phi}p')}\;\int_{\R^3}\frac{d^3p}{2|\vec{p}|}\,\tilde{f}^T
(r^{-1}e^{i\phi}p')p'_M\overline{\tilde{f}(r^{-1}e^{i\phi}p')}\\[10pt]
&=\left(\int_{\R^3}\frac{d^3p}{2|\vec{p}|}\,\tilde{f}^T
(e^{i\phi}p')p'_M\overline{\tilde{f}(e^{i\phi}p')}\right)^2\\[5pt]
&=L(\phi)^2.
\end{align*}

\noindent Thus, if $L$ is zero for some $\phi\in[0,\pi]$, then $F$
is zero on a ray emerging from the origin through $e^{i\phi}$. If
$\phi$ is in $(0,\pi)$, then $F$ is a holomorphic function that is
zero on a set with accumulation points and hence must be zero
entirely. If $\phi=0$ or $\phi=\pi$, then F has zero boundary
values on a set that is open in the boundary and hence must be
zero by the Schwartz reflection principle. So, since $L$ is
nonnegative by definition, either it is zero everywhere or
nowhere.\\

\noindent It remains to show that in the latter case $L$ is
logarithmically convex. Let $\a\in(0,1)$ and
$\C_{+,\a}\doteq\{z\in\C_+\,|\,\arg z<\frac{\pi}{1+\a}\}$.
Consider the function
\begin{align}\label{Gl:DefinitionvonFalpha}
\C_{+\a}\ni z\longmapsto F_{\a}(z)\doteq
\int_{\R^3}\frac{d^3p}{2|\vec{p}|}\,\tilde{f}(z^{1+\a}p')p'_M\overline{\tilde{f}(\bar{z}^{\a-1}p')}\,\in\,\C.
\end{align}

\noindent The integrand is holomorphic, as $z\mapsto z^{1+\a}$ is
on $\C_{+,\a}$. Furthermore, if
$z\in\overline{\C_{+,\a}}\backslash\{0\}$, then
$\overline{z}^{\a-1},\,z^{\a+1}\in\overline{\C_+}\backslash\{0\}$.
So by (\ref{Gl:WichtigeAbschaetzung}) we have:
\begin{align*}
\left|\frac{1}{2|\vec{p}|}\,\tilde{f}(z^{1+\a}p')p'_M\overline{\tilde{f}(\bar{z}^{\a-1}p')}\right|\;\leq\;
\frac{2C_N^2}{(1+|z|^{2\a}|\vec{p}|^2)^N}
\end{align*}

\noindent for all $z\in\overline{\C_{+,\a}}\backslash\{0\}$ and
$\vec{p}\in\R^3$. Therefore the integral exists for all
$z\in\overline{\C_{+,\a}}\backslash\{0\}$. The integrand is
uniformly bounded by an integrable function if $z$ varies in some
compact subset of $\C_{+,\a}$. So $F_{\a}$ is holomorphic on
$\C_{+,\a}$ and has continuous boundary values for $r\in\R^+$
given by
\begin{align*}
\lim_{z\to
r}&\;\int_{\R^3}\frac{d^3p}{2|\vec{p}|}\,\tilde{f}(z^{1+\a}p')p'_M\overline{\tilde{f}(\bar{z}^{\a-1}p')}\;=\;
\int_{\R^3}\frac{d^3p}{2|\vec{p}|}\,\tilde{f}(r^{1+\a}p')p'_M\overline{\tilde{f}(r^{\a-1}p')}\\[5pt]
\;=&\;r^{-3\a}\int_{\R^3}\frac{d^3p}{2|\vec{p}|}\,\tilde{f}(rp')p'_M\overline{\tilde{f}(r^{-1}p')}\;=\;r^{-3\a}F(r).
\end{align*}

\noindent So we see that the two functions $z\mapsto F_{\a}(z)$
and $z\mapsto z^{3\a}F(z)$ are both holomorphic on $\C_{+,\a}$ and
have the same continuous boundary values on $\R^+$. So, by an
application of the Schwartz reflection principle, they have to be
equal:
\begin{align}\label{Gl:Wichtig}
F(z)=z^{-3\a}F_{\a}(z)
\end{align}

\noindent on $\overline{\C_{+,\a}}\backslash\{0\}$. So, for every
$0<\phi<\frac{\pi}{1+\a}$ we have
\begin{align*}
L(\phi)^2&=\,|e^{-3i\a}F_{\a}(e^{i\phi})|^2\\[5pt]
&=\Bigg|\int_{\R^3}\frac{d^3p}{2|\vec{p}|}\,\tilde{f}(e^{i(1+\a)\phi}p')p'_M
\overline{\tilde{f}(e^{i(1-\a)\phi}p')}\Bigg|^2\\[10pt]
&\leq\;\int_{\R^3}\frac{d^3p}{2|\vec{p}|}\,\tilde{f}(e^{i(1+\a)\phi}p')p'_M
\overline{\tilde{f}(e^{i(1+\a)\phi}p')}\;\int_{\R^3}\frac{d^3p}{2|\vec{p}|}\,\tilde{f}(e^{i(1-\a)\phi}p')p'_M
\overline{\tilde{f}(e^{i(1-\a)\phi}p')}\\[10pt]
&=L(\phi(1+\a))L((1-\a)\phi).
\end{align*}

\noindent This means that for every $\phi\in(0,\pi)$ there is a
$\d>0$ such that
\begin{align*}
L(\phi)^2\,\leq\,L(\phi+\e)L(\phi-\e)
\end{align*}

\noindent for all $\e<\d$. Taking the logarithm on both sides, we
get
\begin{align*}
\frac{d^2}{d\phi^2}\ln L(\phi)\,=\,\lim_{\e\to0}\frac{\ln
L(\phi+\e)+\ln L(\phi-\e)-2\ln L(\phi)}{\e^2}\;\geq 0.
\end{align*}

\noindent So $L$ is logarithmically convex, and thus the theorem is proven.\\

\noindent Now we relate $L$ to the twopoint-function of $\BHB$. Let $z=re^{i\phi}\in\overline{\C_+}\backslash\{0\}$.
Then by scaling we have
\begin{align*}
\int_{\R^3}\frac{d^3p}{2|\vec{p}|}\tilde{f}^T(zp')p'_M\overline{\tilde{f}(zp')}
\;=\;\frac{1}{r^3}L(\phi).
\end{align*}

\noindent Now consider the sequence $z_n=1+in\l$ for $\l>0$. Then
$z_n=r_ne^{i\phi_n}$ with $r_n=(\cos\phi_n)^{-1}$. With $L$ as in
(\ref{Gl:DefinitionvonL}) we see that the two series
\begin{align*}
\sum_{n=0}^{\infty}(-1)^n\,\cos^3(\phi_n)L(\phi_n),\qquad\qquad
\sum_{n=1}^{\infty}(-1)^{n-1}\,|\cos^3(\pi-\phi_n)|L(\pi-\phi_n)
\end{align*}

\noindent are absolutely convergent. If we write $L$ in its
explicit integral form (\ref{Gl:DefinitionvonL}) and use
(\ref{Gl:Abschaetzung2}), we may interchange integration and
summation because of dominated convergence, and so we get:
\begin{align}\nonumber
&2\pi\sum_{n=0}^{\infty}(-1)^n\,\cos^3(\phi_n)L(\phi_n)\,+
\,2\pi\sum_{n=1}^{\infty}(-1)^{n-1}\,|\cos^3(\pi-\phi_n)|L(\pi-\phi_n)\\[5pt]\nonumber
=\;&2\pi\,\sum_{n=0}^{\infty}\,(-1)^n\,\int_{\R^3}\frac{d^3p}{2|\vec{p}|}\,\tilde{f}^T\Big((1+i\l
n)p'\Big)p'_M\overline{\tilde{f}\Big((1+i\l n)\Big)}\\\nonumber&\quad+\,
2\pi\,\sum_{n=1}^{\infty}\,(-1)^{n-1}\,\int_{\R^3}\frac{d^3p}{2|\vec{p}|}\,\tilde{f}^T\Big((-1+i\l
n)p'\Big)p'_M\overline{\tilde{f}\Big((-1+i\l n)\Big)}\\[5pt]\nonumber
=\;&(2\pi)^{-3}\int_{\R^3}\frac{d^3p}{2|\vec{p}|}\;\int\,dx\,dy\,f^T(x)p'_M\overline{f(y)}\,
e^{i(p',x-y)}\left(\sum_{n=0}^{\infty}(-e^{-\l(p',x+y)})^n\right)\\[10pt]\nonumber
&\;\;+(2\pi)^{-3}\int_{\R^3}\frac{d^3p}{2|\vec{p}|}\;\int\,dx\,dy\,f^T(x)p'_M\overline{f(y)}\,
e^{-i(p',x-y)}\left((e^{-\l(p',x+y)})\sum_{n=1}^{\infty}(-e^{-\l(p',x+y)})^{(n-1)}\right)\\[10pt]\nonumber
=\;&(2\pi)^{-3}\int
dp\,\d(p^2)\e(p_0)\;\int\,dx\,dy\,f^T(x)p_M\overline{f(y)}\,
\frac{e^{i(p,x-y)}}{1+e^{\l(p,x+y)}}\\[10pt]\label{Gleichungeins}
=\;&\BHB\big(\bpsi(\bar{f})\psi(f)\big).
\end{align}

\noindent By making use of the anticommutation relations, one also
gets
\begin{align}\nonumber
\BHB\big(\psi(f)&\bpsi(\bar{f})\big)\;\\\label{Gleichungzwei}&=\;2\pi\sum_{n=1}^{\infty}(-1)^{n-1}\,\cos^3(\phi_n)L(\phi_n)\,+
\,2\pi\sum_{n=0}^{\infty}(-1)^{n}\,|\cos^3(\pi-\phi_n)|L(\pi-\phi_n).
\end{align}

\noindent So in order to check whether $\BHB$ is a state, we have
to check whether the two series described above are nonnegative
for every choice of $f\in\D(V^+,\C^2)$. This will be done in the
following.

\begin{Theorem}\label{Satzeins}
Let $L:[0,\pi]\to\R^+$ be a continuous, convex function that is
smooth on $(0,\pi)$. Define $r:[0,\pi]\to\R$ by
$r(\phi)=|\cos^3\phi|$ and $g(\phi)=r(\phi)L(\phi)$. Let
furthermore $\{\phi_n\}_{n\in\N}$ be a monotonically increasing
sequence in $[0,\frac{\pi}{2})$ converging to $\frac{\pi}{2}$,
such that
\begin{align}\label{Gl:Absconvergence}
\sum_{n=0}^{\infty}g(\phi_n)\,<\,\infty.%,\qquad\qquad\sum_{n=0}^{\infty}g(\pi-\phi_n)\,<\,\infty.
\end{align}

\noindent Then the two series
\begin{align}\label{Gl:SeriesA}
A_L&\doteq \sum_{n=0}^{\infty}(-1)^n \Big[g(\phi_n) +
g(\pi-\phi_{n+1})\Big]\\[5pt]\label{Gl:SeriesB}
B_L&\doteq \sum_{n=0}^{\infty}(-1)^n \Big[g(\pi-\phi_n) +
g(\phi_{n+1})\Big]
\end{align}

\noindent converge absolutely and are both nonnegative.
\end{Theorem}

\noindent\textbf{Proof:} Because $L$ is continuous at
$\phi=\frac{\pi}{2}$, it follows from the convergence of
(\ref{Gl:Absconvergence}) that $\sum_ng(\pi-\phi_n)$ converges,
too. Since $L>0$ we have $g\geq0$, and therefore the two series
(\ref{Gl:SeriesA}) and (\ref{Gl:SeriesB}) converge absolutely.\\

\noindent To establish positivity of the two series, we first show
that the function $g$ is either monotonous on $[0,\frac{\pi}{2}]$
or on $[\frac{\pi}{2},\pi]$: Assume $g$ not to be monotonous on
$[0,\frac{\pi}{2}]$. Then there is a
$\phi_{N}\in(0,\frac{\pi}{2})$ with $g'(\phi_N)=0$. Since $L>0$
and $r'(\phi_N)<0$, we then have that
\begin{align*}
L'(\phi_N)=\frac{-r'(\phi_N)L(\phi_N)}{r(\phi_N)}\,>\,0,
\end{align*}

\noindent and thus, since $L$ is convex, $L'>0$ on
$[\frac{\pi}{2},\pi)$. Therefore, for all
$\phi\in[\frac{\pi}{2},\pi)$, we have that
\begin{align*}
g'(\phi)=r(\phi)L'(\phi)+r'(\phi)L(\phi)\,>\,0,
\end{align*}

\noindent since $r$ and $r'$ are non-negative on
$[\frac{\pi}{2},\pi)$. So $g$ is monotonous on
$[\frac{\pi}{2},\pi]$.\\

\noindent Now assume $g$ to be not monotonous on
$[\frac{\pi}{2},\pi]$. Replace $L$ by $\overline{L}$ given by
\begin{align}\label{Gl:Lquer}
\overline{L}(\phi)\doteq L(\pi-\phi).
\end{align}

\noindent The function $\overline{L}$ is convex, too, and
$\overline{g}=r\cdot\overline{L}$ is not monotonous on
$[0,\frac{\pi}{2}]$. Thus, the above argument can be applied to
$\overline{g}$ instead of $g$ and shows that $g$ is monotonous on
$[0,\frac{\pi}{2}]$.\\

\noindent Since $g(0)>0<g(\pi)$ and $g(\frac{\pi}{2})=0$, we know
that either $g$ is monotonically decreasing on $[0,\frac{\pi}{2}]$
or monotonically increasing on $[\frac{\pi}{2},\pi]$ (or both).
Without loss of generality, we can assume the latter to be the
case. Otherwise we could replace $L$ by $\overline{L}$ as in
(\ref{Gl:Lquer}), since by (\ref{Gl:SeriesA}) and
(\ref{Gl:SeriesB}) we see that $A_L=B_{\overline{L}}$ and
$B_L=A_{\overline{L}}$. So by this replacement both series are
just interchanged.\\

\noindent Thus, from now on, $g$ will be monotonically increasing
on $[\frac{\pi}{2},\pi]$. There are two possibilities: $L$ may or
may not be monotonous on $[0,\frac{\pi}{2}]$.
\begin{itemize}
\item $L$ is monotonous on $[0,\frac{\pi}{2}]$:\\

\noindent Let $L$ be monotonically decreasing on
$[0,\frac{\pi}{2}]$, then $g$ is, too. This means that $g$ is
monotonous on $[0,\frac{\pi}{2}]$ and $[\frac{\pi}{2},\pi]$. By
reordering of (\ref{Gl:SeriesA}) and (\ref{Gl:SeriesB}), we get:
\begin{align}
A_L=\sum_{n=0}^{\infty}
\Big[g(\phi_{2n})-g(\phi_{2n+1})\Big]\;+\;\sum_{n=1}^{\infty}
\Big[g(\pi-\phi_{2n-1})-g(\pi-\phi_{2n})\Big]\\[5pt]
B_L=\sum_{n=0}^{\infty}
\Big[g(\pi-\phi_{2n})-g(\pi-\phi_{2n+1})\Big]\;+\;\sum_{n=1}^{\infty}
\Big[g(\phi_{2n-1})-g(\phi_{2n})\Big].
\end{align}

\noindent Since $\phi_m\leq\phi_{m+1}$ for all $m$, we see that
every expression in square brackets is non-negative, and so are
$A_L$ and $B_L$.\\

\noindent Let $L$ be monotonically increasing on
$[0,\frac{\pi}{2}]$. Thus, since $L''>0$, we have, for all
$\phi\in(0,\frac{\pi}{2})$ that $0<L(\phi)<L(\pi-\phi)$ and
$0<L'(\phi)<L'(\pi-\phi)$. So, for such a $\phi$ we have
\begin{align*}
|g'(\phi)|\,&\leq\,|r'(\phi)|\cdot|L(\phi)|+|r(\phi)|\cdot|L'(\phi)|\\[5pt]
&\leq\,r'(\pi-\phi)L(\pi-\phi)+r(\pi-\phi)L'(\pi-\phi)\\[5pt]
&=g'(\pi-\phi).
\end{align*}

\noindent Thus, for $0\leq a\leq b\leq \frac{\pi}{2}$ we have
\begin{align}\label{Gl:Abschaetzung1}
|g(a)-g(b)|\,\leq\,\int_a^b|g'(\phi)|d\phi\,\leq\,\int_a^bg'(\pi-\phi)d\phi=g(\pi-a)-g(\pi-b).
\end{align}

\noindent We rewrite (\ref{Gl:SeriesA}) and (\ref{Gl:SeriesB}) as
follows:
\begin{align*}
A_L&=g(\phi_0)\,+\,\sum_{n=1}^{\infty}\Big[g(\pi-\phi_{2n-1})-g(\pi-\phi_{2n})+g(\phi_{2n})-g(\phi_{2n-1})\Big]\\[5pt]
B_L&=g(\phi_0)\,+\,\sum_{n=0}^{\infty}\Big[g(\pi-\phi_{2n})-g(\pi-\phi_{2n+1})+g(\phi_{2n+1})-g(\phi_{2n})\Big].
\end{align*}

\noindent By (\ref{Gl:Abschaetzung1}) and $\phi_n\leq\phi_{n+1}$
for all $n\in\N$, the expressions in square brackets are
non-negative for all $n\in\N$, and since $g$ is positive, both
series are positive as well.

\item $L$ is not monotonous on $[0,\frac{\pi}{2}]$:\\

\noindent Since $L$ is convex, there is a
$\phi_{Null}\in(0,\frac{\pi}{2})$ such that $L$ is monotonically
decreasing on $[0,\phi_{Null}]$ and monotonically increasing on
$[\phi_{Null},\frac{\pi}{2}]$. So $|g'(\phi)|\leq g'(\pi-\phi)$
for all $\phi\in[\phi_{Null}, \frac{\pi}{2}]$, by the same
argument as above. Thus, relation (\ref{Gl:Abschaetzung1}) is
valid for all $\phi_{Null}\leq a\leq b\leq \frac{\pi}{2}$. This
means that for $\phi_0$ such that $\phi_{Null}\leq\phi_0$ we are
done. If $\phi_0<\phi_{Null}$, there is $p\in\N$ such that
$\phi_p\leq\phi_{Null}\leq\phi_{p+1}$. Now consider the sequence
$\{\tilde{\phi}\}_{n\in\N}$, which is given by
\begin{align*}
&\tilde{\phi}_n=\phi_n\qquad\mbox{for }n\leq p\\[5pt]
&\tilde{\phi}_{p+1}=\tilde{\phi}_{p+2}=\phi_{Null}\\[5pt]
&\tilde{\phi}_{n+3}=\phi_{n+1}\qquad\mbox{for }n\geq p.
\end{align*}

\noindent One easily sees by (\ref{Gl:SeriesA}) and
(\ref{Gl:SeriesB}) that $A_L$ and $B_L$ evaluated with the
sequence $\{\tilde{\phi}\}_n$ have the same values as evaluated
with the sequence $\{\phi\}_n$. So, without loss of generality, we
may assume $\phi_{Null}$ to be a member of $\{\phi_n\}_{n\in\N}$.
By what we just said, we may also assume that
$\phi_{Null}=\phi_{2m+1}=\phi_{2m}$ for some $m\in\N$. Again, we
reorder the series (\ref{Gl:SeriesA}) and (\ref{Gl:SeriesB}) and
get:
\begin{align*}
A_L=\sum_{n=0}^{m-1}\Big[g(\phi_{2n})-g(\phi_{2n+1})\Big]\,+\,g(\phi_{2m})\,+\,\sum_{n=m+1}^{\infty}
\Big[g(\phi_{2n})-g(\phi_{2n-1})\Big]&\\[5pt]
+\sum_{n=1}^{m}\Big[g(\pi-\phi_{2n-1})-g(\pi-\phi_{2n})\Big]\;+
\;\sum_{n=m+1}^{\infty}\Big[g(\pi-\phi_{2n-1})-g(\pi-\phi_{2n})\Big]&.
\end{align*}

\noindent Since $L$ is monotonically decreasing on
$[0,\phi_{2m}]$, so is $g$. Thus, the first sum is nonnegative.
$L$ is increasing on $[\phi_{2m},\frac{\pi}{2}]$ and therefore
relation (\ref{Gl:Abschaetzung1}) is valid for all $\phi_{2m}\leq
a\leq b\leq \frac{\pi}{2}$. So the second sum could be negative,
but the fourth sum dominates it, so the sum of both is
nonnegative. That the third sum is nonnegative is a consequence of
the fact that $g$ is monotonically increasing on $[\pi-\phi_{2m},
\pi]$. So $A_L$ is nonnegative. Similarly we have:
\begin{align*}
B_L=\sum_{n=1}^{m}\Big[g(\phi_{2n-1})-g(\phi_{2n})\Big]\,+\,g(\phi_{2m+1})\,+\,\sum_{n=m+1}^{\infty}
\Big[g(\phi_{2n+1})-g(\phi_{2n})\Big]&\\[5pt]
+\sum_{n=0}^{m}\Big[g(\pi-\phi_{2n})-g(\pi-\phi_{2n+1})\Big]\;+
\;\sum_{n=m+1}^{\infty}\Big[g(\pi-\phi_{2n})-g(\pi-\phi_{2n+1})\Big]&.
\end{align*}

\noindent Again, $L$ is monotonically decreasing on
$[0,\phi_{2m+1}]$ and monotonically increasing on
$[\phi_{2m+1},\frac{\pi}{2}]$. So, by analogous arguments as
above, $B_L$ is nonnegative, too.
\end{itemize}
\noindent This completes the proof of the theorem.\\[10pt]

\noindent So, by theorems \ref{Satzzwei} and \ref{Satzeins} we have shown that the expressions in (\ref{Gleichungeins}) 
and (\ref{Gleichungzwei}) are nonnegative, and this means that (\ref{Gl:BHB}) defines a
gauge-invariant, quasifree state on $\F(V^+)$. This is a closed
sub-$C^*$-algebra of $\F$, and hence we can extend $\BHB$ to all
of $\F$. Since due to the anticommutation relations positivity is
equivalent to boundedness by one, the theorem of Hahn-Banach
guarantees that such extension can be chosen to be positive and
hence to be a state, too. The explicit form of such an extension
will not be in our interest in the following.

\section{The physical picture}

\noindent A straightforward calculation shows that
\begin{align}\label{Gl:Thermal1}
\BHB\big(\l^{\mm\n}(x)\big)\,=\,\o_{2\l x}\big(\l^{\mm\n}(x)\big),
\end{align}

\noindent where $\o_{2\l x}$ is the KMS-state for $\b=2\l x$.
Since the square of the temperature $\b\mapsto(\b,\b)^{-1}=T^2$
solves the wave equation on $V^+$, $T^2$ is an admissible
macroobservable that can be approximated by elements in $\S_x$
like in (\ref{Gl:approx}). Its expectation value in $\BHB$ is
\begin{align*}
\BHB(T^2)(x)\,=\,\frac{1}{4\l^2(x,x)}.
\end{align*}

\noindent So the temperature diverges on the apex and the boundary
of the forward lightcone $V^+$. The phase-space density in the
state $\BHB$ is given by
\begin{align*}
\BHB(N_p)(x)\,=\,\frac{(2\pi)^{-3}}{1+e^{2\l(x,p)}}.
\end{align*}

\noindent This shows that at a point near $x$ the boundary of
$V^+$, the dominant contribution of the particle density is made
by the particles with momentum $p$ proportional to $x$. If $x$
approaches the boundary, the total number of particles increases
and their speed tend to the speed of light. Thus, one would
interpret this state as the result of a Hot Bang, after which a
vast number of particles emanates into space with speed of light.
Furthermore, for each observer 'inside the shockwave' the
temperature of the system decreases with time as $T=(2\l t)^{-1}$.
On can show that for each timelike vector $a\in V^+$ we have
\begin{align*}
\lim_{t\to\infty}\;\BHB\big(\a_{ta}(A)\big)\;=\;\o_{\infty}\big(A\big),
\end{align*}

\noindent so the state $\BHB$ approaches the vacuum in timelike
infinity. Thus, the name Hot Bang-state is appropriate.

\section{Less thermal observables}

\noindent Let us now come to a mathematical subtlety of this approach of characterizing local equilibrium states.
The choice of thermal observables (\ref{Gl:Lambdas1}) is
not unique. In \cite{BOR02} a method is shown how to build a space
of thermal observables $\T_x$ for generic models. In our case
$\S_x$ is a proper subspace of $\T_x$. In the analysis, one could
have chosen a larger or smaller subspace of $\T_x$ as space of thermal observables,
thus admitting less or more states to be locally close to
equilibrium in the sense of the definition \ref{Def:Esso}. This is
a potent method to impose a hierarchy onto states, by ordering
them in terms of local closeness to reference states.

\noindent One can see from (\ref{Gl:Lambdas1}) and (\ref{FormL}) that the
thermal observables $\l^{\mm\n}(x)$ are symmetric in the first
$m=\deg\mm$ indices, but the last one has to be treated
separately. The corresponding thermal functions $\b\mapsto
L^{\mm\n}(\b)$, though, are symmetric in all $m+1$ indices. This
indicates that there are thermal observables that are not zero by
themselves but vanish in all reference states, for example
$\l^{\m\n}(x)-\l^{\n\m}(x)$. Hence, in the sense of the
approximation (\ref{Gl:approx}) the $\l^{\mm\n}(x)$ contain
redundancies. Therefore one could symmetrize the $\l^{\mm\n}(x)$
\begin{align}\label{Gl:Tilde}
\tilde{\l}^{\mm\n}(x)\,\doteq\,\sum_{\pi\in
P_{m+1}}\,\l^{\pi(\mm\n)}(x).
\end{align}

\noindent in order to make the series used in (\ref{Gl:approx})
less ambiguous. For these new thermal observables one does not
need to distinguish between different indices and is able to write
$\tilde{\l}^{\mm}(x)$ instead. We denote the space of the
$\tilde{\l}^{\mm}(x)$ by $\S'_x$. So $\S'_x$ is a proper subspace
of $\S_x$ for every $x$, thus there are potentially more
$\S'_{\O}$-thermal than $\S_{\O}$-thermal states. The thermal
functions induced by the $\tilde{\l}^{\mm}(x)$ are the same as the
ones induced by the $\l^{\mm\n}(x)$, of course, hence all the
$\S'_{\O}$-thermal states can be continued to the admissible
macro-observables, too.\\

\noindent One can show \cite{BAB04} that the Weyl equations
(\ref{Diraczwei}) induce differential equations on the
$\l^{\mm\n}(x)$, namely the following:

\begin{align}\label{Gl:LambdaBGL3}
\del_{\t}\l^{\t\mm\n}(x)&=0\\[5pt]\label{Gl:LambdaBGL4}
\Box\l^{\mm\n}(x)+\l_{\t}{}^{\t\mm\n}(x)&=0\\[5pt]
\label{Gl:LAMBDA2}
 \s^{\n,\dr s}\;\s_{\t,\,s\dt}\l_{\n}{}^{\mm\t}(x)-\s^{\n,\dr
 s}\;\s_{\t,\,s\dt}\,\del_{\n}\l^{\mm\t}(x)&=0\\[5pt]\label{Gl:LAMBDA1}
 \s^{\n,\dr s}\;\s_{\t,\,t\dr}\l_{\n}{}^{\mm\t}(x)+\s^{\n,\dr
s}\;\s_{\t,\,t\dr}\,\del_{\n}\l^{\mm\t}(x)&=0.
\end{align}

\noindent These differential equations have implications for the
space-time behavior of the expectation values of thermal
macroobservables $\Xi$ in $\S_{\O}$-thermal states $\o$:

\begin{align}\label{KrasseDGL1}
\Box\,\o(\Xi)(x)&=0\\[5pt]\label{KrasseDGL2}
\del^{\m}\,\o(\del_{\m}\Xi)(x)&=0\\[5pt]\label{KrasseDGL3}
\del_{\m}\o(\del_{\n}\Xi)(x)-\del_{\n}\o(\del_{\m}\Xi)(x)&=0.
\end{align}

\noindent Since the $\tilde{\l}^{\mm}(x)$ are less than the
$\l^{\mm\n}(x)$, not all of the equations (\ref{KrasseDGL1}) -
(\ref{KrasseDGL3}) have to hold for $\S'_{\O}$-thermal states. One
can say something at least. If one takes the trace over the free
spinor indices in (\ref{Gl:LAMBDA2}) and (\ref{Gl:LAMBDA1}), one
obtains $\l_{\n}{}^{\mm\n}(x)=0$ and $\del_{\n}\l^{\mm\n}(x)=0$.
Let $\eta$ be the Minkowski metric, then we get

\begin{align*}
\eta_{\t\rho}\,\sum_{\pi\in
P_{m+3}}\l^{\pi(\rho\t\mm\n)}(x)&\;=\;\eta_{\t\rho}\sum_{\begin{array}{c}\scriptstyle\pi\in
P_{m+3}\\[-3pt]\scriptscriptstyle\pi(1)\neq
m+3\\[-3pt]\scriptscriptstyle\pi(2)\neq
m+3\end{array}}\l^{\pi(\rho\t\mm\n)}(x)\\[10pt]
&\;=\;\eta_{\t\rho}(m+2)(m+1)\sum_{\pi\in
P_{m+1}}\l^{\t\rho\pi(\mm\n)}(x),
\end{align*}

\noindent since the $\l^{\mm\n}(x)$ are symmetric in all but the
last index. So we have because of (\ref{KrasseDGL2}):
\begin{align*}
0&\;=\;(m+2)(m+1)\sum_{\pi\in
P_{m+1}}\Box\l^{\pi(\mm\n)}+(m+2)(m+1)\eta_{\t\rho}\sum_{\pi\in
P_{m+1}}\l^{\t\rho\pi(\mm\n)}(x)\\[10pt]
&\;=\;(m+2)(m+1)\sum_{\pi\in
P_{m+1}}\Box\l^{\pi(\mm\n)}+\eta_{\t\rho}\sum_{\pi\in
P_{m+3}}\l^{\pi(\t\rho\mm\n)}(x),
\end{align*}

\noindent and hence
\begin{align}\label{Gl:TildeGL1}
\frac{1}{(m-1)!}\Box\tilde{\l}^{\mm}(x)+\frac{1}{(m+1)!}\tilde{\l}_{\t}{}^{\t\mm}(x)\;=\;0.
\end{align}

\noindent With $\del_{\n}\l^{\mm\n}(x)=0$, on the other hand, it
is easy to show that
\begin{align}\label{Gl:TildeGL2}
\del_{\n}\tilde{\l}^{\n\mm}(x)\;=\;0.
\end{align}

\noindent Equations (\ref{Gl:TildeGL1}) and (\ref{Gl:TildeGL2})
suffice to show that the equations
\begin{align}\label{Gl:TildeEQN}
\Box
\o(\Xi)(x)\,=\,0,\qquad\qquad\del_{\m}\o(\del^{\m}\Xi)(x)\,=\,0
\end{align}

\noindent hold for $\S'_{\O}$-thermal states $\o$, too. On the
other hand, no equations of the form
\begin{align*}
A \s^{\n,\dr
s}\;\s_{\t,\,s\dt}\tilde{\l}_{\n}{}^{\mm\t}(x)-B\s^{\n,\dr
 s}\;\s_{\t,\,s\dt}\,\del_{\n}\tilde{\l}^{\mm\t}(x)&=0\\[5pt]
A \s^{\n,\dr
s}\;\s_{\t,\,t\dr}\tilde{\l}_{\n}{}^{\mm\t}(x)+B\s^{\n,\dr
s}\;\s_{\t,\,t\dr}\,\del_{\n}\tilde{\l}^{\mm\t}(x)&=0
\end{align*}

\noindent for constants $A,B$ can hold. Otherwise, by taking the
trace over the free spinor indices on both equations one would
obtain $\tilde{\l}_{\n}{}^{\n\mm}(x)=0$, violating equation
(\ref{Gl:TildeGL1}). Nevertheless, the two equations
(\ref{Gl:TildeEQN}) are enough to establish the main results of
the bosonic case: The phase-space density $\o(N_p)(x)$ of
$\S'_{\O}$-thermal states satisfies the massless, free Boltzmann
equation, and all statements about the breaking of time-reversal symmetry carry over to
the fermionic case. So, in the analysis, one could look for
$\S'_{\O}$-thermal states instead of $\S_{\O}$-thermal states. By
restricting oneself to less thermal observables, one could find
more states of interest that still exhibit interesting properties
one would expect of local equilibrium states.\\

\noindent\textbf{\Large{Acknowledgements}}\\

\noindent I would like to thank Professor Detlev Buchholz for
fruitful discussions, his patience and time. Furthermore I am in
debt to all members of the LQP-group in the Institute for Theoretical Physics in
G\"ottingen for their time and support. Furthermore I would like to thank Thomas Thiemann for his help.

\end{document}